\documentclass[aps,prb,twocolumn,amsmath,superscriptaddress]{revtex4-2}
\usepackage[caption=false]{subfig}
\usepackage{xcolor,graphicx} 
\usepackage{MnSymbol} 
\usepackage{mathtools}
\usepackage{braket} 
\usepackage{tikz} 

\usepackage[utf8]{inputenc} 
\usepackage{hyperref}
\usepackage{mhchem}

\begin{document}

\title{Promising regimes for the observation of topological degeneracy in spin chains}

\author{Alexander Sattler}
\author{Maria Daghofer}
\affiliation{Institut f\"ur Funktionelle Materie und Quantentechnologien,
Universit\"at Stuttgart,
70550 Stuttgart,
Germany}

\date{\today}

\begin{abstract}
Both the Haldane spin chain and a topologically dimerized chain feature topologically protected edge states that are expected to be robust against some kind of noise. To elucidate whether it might be feasible to create such edge states in dimerized chains in a controlled manner in solid states environments, e.g. as spin chains on surfaces, as has already been successfully achieved with the Haldane chain, we investigate their robustness with respect to long-range coupling, anisotropies and finite chain length. The theoretical investigation is based on an alternating Heisenberg spin chain with spin-$1/2$, which is investigated using exact diagonalization. We find that dimerized chains and Haldane chains have robustness against long-range coupling and anisotropies. In particular, dimerized spin chains are significantly more robust than Haldane chains.
\end{abstract}

\maketitle

\section{Introduction} \label{sec:introduction}
Topological order and symmetry-protected topological order and their signatures are currently a topic of considerable
research  interest~\cite{RevModPhys.89.040502,RevModPhys.89.041004,RevModPhys.88.035005,RevModPhys.83.1057,RevModPhys.82.3045}, both in non-interacting and in interacting systems, and in
one as well as in higher dimensions. A state with topological order can only
go over into a topologically trivial one by closing a gap, and surface, edge,
or end states are thus an important feature of topologically nontrivial
states. Considerable effort is thus spent on realizing and investigating such
states, whether in topological states emerging in solids or engineered
from other building blocks.

In addition to providing signatures of topological order, surface, edge, and end
states are also sought out as a potential use of topological systems. For
instance, chiral edges are protected against
backscattering~\cite{RevModPhys.82.3045,RevModPhys.83.1057}, and helical edge states transport only one spin in each
direction~\cite{RevModPhys.82.3045,RevModPhys.83.1057}, with uses in spintronics~\cite{doi:10.1126/science.1065389,HIROHATA2020166711,RevModPhys.76.323,doi:10.1146/annurev-conmatphys-070909-104123}. Topological protection has been proposed as a transport
channel for quantum information~\cite{2017,edge_quant_commun}, and excitations of fractional quantum Hall states can be used in
quantum computation~\cite{RevModPhys.80.1083}. End states of topological spin chains form effective spin-$1/2$
degrees of freedom and have been conjectured to provide some
protection from decoherence~\cite{DELGADO201740} and have been proposed as a
building block for quantum computation~\cite{PhysRevLett.105.040501}.

The Haldane
phase~\cite{RevModPhys.89.040502,RevModPhys.89.041004,I_Affleck_1989,inbook_regnault,PhysRevB.85.075125},
a symmetry-protected topological phase based on a conjecture of
Haldane~\cite{PhysRevA.93.464,PhysRevLett.50.1153}, is an early
one-dimensional example  for an interaction-based topological scenario. In this 
chain with spin one, superexchange induces an antiferromagnetic (AFM)
coupling between the spins. The  resulting AFM ground state is
separated from the lowest excitations by a gap, which can be measured by neutron
scattering, susceptibility measurements and magnetization
measurement~\cite{PhysRevB.54.R6827,PhysRevB.39.4820,10.1063/1.340736,PhysRevLett.56.371,J_P_Renard_1987,PhysRevLett.63.86,PhysRevB.38.543,REGNAULT199571,DARRIET1993409,PhysRevLett.63.1424,Cizmar_2008,PhysRevLett.90.087202}.

\begin{figure}
  \includegraphics[width=0.7\columnwidth]{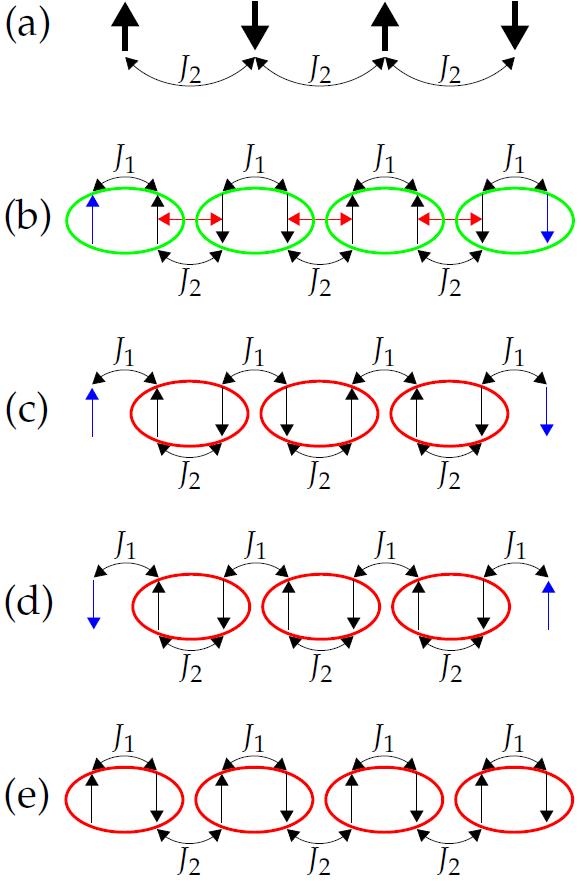}
    \caption{Sketches for the ground states of a chain of spin-$1/2$ with alternating Heisenberg exchange couplings $J_1$ and $J_2$. 
      In the sketch for the Haldane phase (b), the green ellipses are the effective spin-$1$, 
      the red arrows indicate their AFM coupling resulting in singlets, and the blue arrows are the free edge spin-$1/2$.
    The sketch (a) shows the effective AFM spin-$1$ chain.
 The spin-$1$ in the sketch (a) are the green ellipses in the sketch (b).
      In the sketches for the dimer phases (c), (d) and (e), the red ellipses
      are singlets,  and 
      the blue arrows are the free edge spins with
      spin-$1/2$.
\label{fig:cartoon}}
\end{figure}

Conceptually, the properties of the Haldane chain can be understood by
splitting each spin-$1$ into two spin-$1/2$  and then coupling those into
singlets~\cite{PhysRevB.85.075125,inbook_regnault,I_Affleck_1989,RevModPhys.89.041004,PhysRevB.48.9555,PhysRevB.46.8268,PhysRevB.45.2207,PhysRevB.46.3486,PhysRevLett.65.3181,1904.02102}, see
Fig.~\ref{fig:cartoon}(a) and Fig.~\ref{fig:cartoon}(b).
One consequence of this singlet
decomposition are the left-over spin-$1/2$ at the edges, also seen in Fig.~\ref{fig:cartoon}(b). These are the
topologically protected edge states, which leave
measurable signature in e.g. electron spin
resonance~\cite{PhysRevLett.65.3181,PhysRevLett.67.1614,doi:10.1143/JPSJ.67.2514,Cizmar_2008,PhysRevLett.95.117202}, nuclear magnetic resonance~\cite{PhysRevLett.83.412} and inelastic neutron scattering~\cite{PhysRevLett.90.087202}. 

Edge states of Haldane chains were theoretically concluded to be also 
accessible to a scanning tunneling microscope
(STM)~\cite{PhysRevLett.111.167201}  and have been successfully measured~\cite{Mishra2021}.
There are several reasons why the ideal
scenario can only be approximated: The two edges of finite
chains, which can be realized in an STM, are coupled, with the coupling
between them suppressed exponentially for longer chains. Moreover, magnetic superexchange between spins is not
strictly restricted to nearest neighbors and longer-ranged interactions in general increase coupling
between the edge states. While they can, in some cases, even be used to decouple
edge states from each other on finite chains, this requires delicate fine
tuning~\cite{PhysRevB.96.060409,PhysRevB.97.174434}. Finally, signatures of
the edge states  are rather sensitive to the $z$-axis
anisotropy~\cite{PhysRevLett.111.167201}.

A second well-known one-dimensional topological state is the
Su-Schrieffer-Heeger (SSH) model~\cite{PhysRevLett.42.1698}, where bonds
alternate between stronger and weaker coupling. It was originally considered for
non-interacting fermions at half filling, where the band structure likewise
has zero-energy edge states if the bonds at the ends are weak ones. These can
be empty or occupied, i.e. act like a spin-$1/2$ degree of freedom. For
non-interacting bosons or finite onsite interactions, these zero-energy
states likewise exist, but next-nearest-neighbor (NNN)
hopping moves them away from zero energy~\cite{PhysRevLett.110.260405}. If
onsite interactions are infinite, i.e. for hard-core bosons that are in one
dimension largely equivalent to fermions~\cite{JordanWigner}, the impact of  NNN
hoppings onto the coupling of the edge states~\cite{SSH_Rydberg_19} is
largely removed. This has
recently enabled the observation of the corresponding edge states in a
dimerized Rydberg-atom chain~\cite{SSH_Rydberg_19}.

Here, we investigate edge states of dimerized~\cite{doi:10.1143/JPSJ.62.3357,Bahovadinov_2019,PhysRevB.46.8268,PhysRevB.46.3486,10.1088/0953-8984/26/27/276002,doi:10.1143/JPSJ.58.4367,Haghshenas_2014,PhysRevB.46.3486,PhysRevB.45.2207,PhysRevB.48.9555,10.7566/JPSJ.85.124712,1508.06129,10.1088/0953-8984/27/16/165602,PhysRevB.87.054402,1212.6012,1904.02102,PhysRev.165.647,PhysRevB.46.3486,PhysRevB.46.8268,PhysRevB.45.2207,PhysRevB.63.144428} spin chains as sketched in Fig.~\ref{fig:cartoon}. For a strong FM
coupling within dimers, the situation  becomes clearly similar to Haldane's
spin-$1$
chain, see
Fig.~\ref{fig:cartoon}(a-b). Indeed,
edge modes whose coupling vanishes exponentially with system size have accordingly been
investigated~\cite{Maruyama_2010}, and the impact of NNN 
couplings on the topological phase diagram has likewise been addressed~\cite{1212.6012,10.1088/0953-8984/26/27/276002,1508.06129,10.7566/JPSJ.85.124712,1904.02102}.
Strong AFM coupling within
as well as between dimers leads to a scenario more similar to the SSH chain~\cite{Bahovadinov_2019},
sketched in Fig.~\ref{fig:cartoon}(d,e). 
The main theoretical difference to the hard-core bosonic chain is that the
latter is equivalent to $X$-$Y$ spins, while the spin model can include coupling of $Z$
components.~\footnote{Alternatively, mapping isotropic Heisenberg exchange onto bosons
generates intersite density-density interactions in addition to the onsite
hard-core constraint.} As a consequence, positive and negative couplings can
no longer be mapped onto each other in the spin model. 

Two perfect spin-$1/2$ edge states imply a four-fold degenerate ground state, as each spin
can be flipped without penalty. 
To be observable, these four states need to be separated from the rest of the spectrum by a gap. 
While perfect degeneracy cannot be expected for finite chains~\cite{T_Kennedy_1990,PhysRevB.50.6277,PhysRevB.48.913}, splitting
within the lowest four states should be much smaller than the gap separating
them from the rest of the spectrum.
Since 
we are here motivated by spin chains in a scanning electron
microscope, AFM superexchange couplings are  easier to achieve
than FM  ones. Moreover, the direction perpendicular to the surface is clearly special,  so that some $z$-axis anisotropy is to be expected~\cite{GAUYACQ201263,doi:10.1126/science.1125398,DELGADO201740,doi:10.1126/science.1082857,doi:10.1126/science.1214131,doi:10.1126/science.1146110,PhysRevLett.111.127203,PhysRevLett.102.256802,doi:10.1021/acs.nanolett.7b02882}. In addition to
idealized dimerized chains, we thus also include anisotropies as well as
NNN couplings and  investigate how they affect
ground state quasi-degeneracy.

The paper is structured as follows:
In Sec.~\ref{section:model} we discuss the Hamiltonian of our model.
In addition, we introduce our method to distinguish between the topological and the topologically trivial phases.
In Sec.~\ref{sec:results} we discuss the results obtained by varying the different parameters in the Hamiltonian introduced in Sec.~\ref{section:model}.
First, in Sec.~\ref{subsec:simple_chain}, we discuss the simplest possible chains.
These chains are isotropic and interactions are limited to the nearest neighbor. 
We show how to use the method, see Sec. \ref{section:model}, to analyze the topological phase transitions.
Therefore, we analyze the different energy gaps between the lowest energies.
In the subsequent chapter, the different parameters of the Hamiltonian are analyzed.
We start with the NNN coupling in Sec.~\ref{subsec:J_NNN}, investigate
$z$-axis anisotropy in Sec.~\ref{subsec:z_anis}, and combine both in Sec.~\ref{sec:z_anis_NNN}.
The inelastic tunneling current which is accessible in an STM is discussed in Sec. \ref{sec:current}.
Section~\ref{sec:discuss} concludes the paper and gives a summary of the results obtained as well as an outlook to promising future studies.
 
\section{Model and Methods}\label{section:model}

\begin{figure}
  \centering
  \includegraphics[width=0.7\columnwidth]{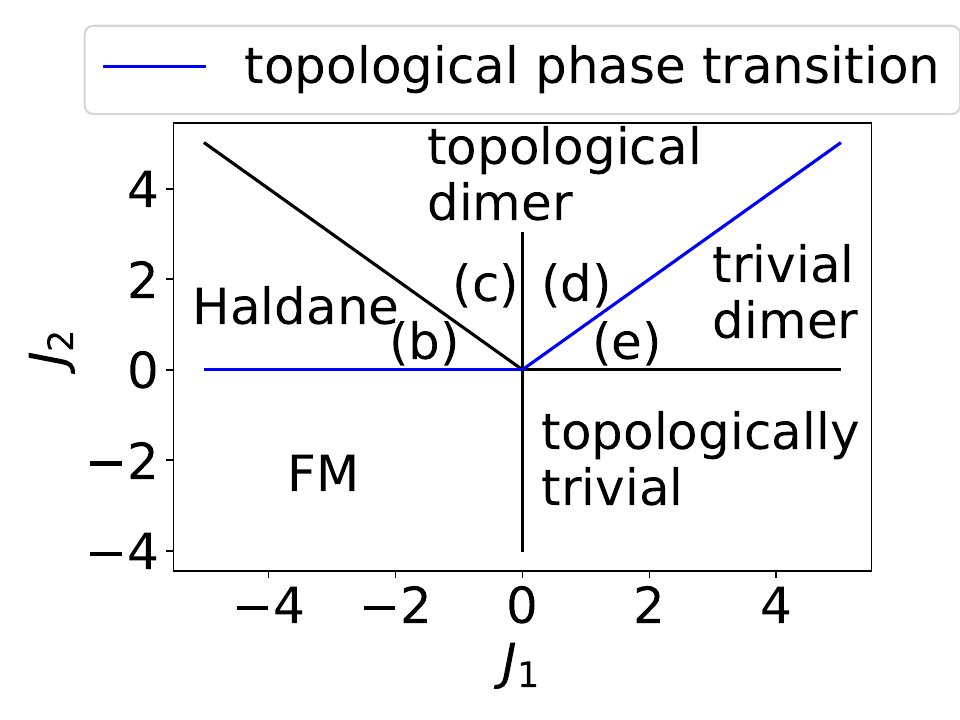}
    \caption{Phase diagram of an infinite alternating Heisenberg spin chain with the alternating coupling constants $J_1$ and $J_2$.
	The blue lines show the topological phase transition. 
        Labels (b)-(e) refer to the sketches shown in Fig.~\ref{fig:cartoon},
        which schematically illustrate the ground states.
    \label{fig:phases_ideal}}
\end{figure}

We start from an idealized dimerized~\cite{doi:10.1143/JPSJ.62.3357,Bahovadinov_2019,Haghshenas_2014,PhysRevB.46.8268,PhysRevB.46.3486,10.1088/0953-8984/26/27/276002,doi:10.1143/JPSJ.58.4367,PhysRevB.46.3486,PhysRevB.45.2207,PhysRevB.48.9555,10.7566/JPSJ.85.124712,1508.06129,10.1088/0953-8984/27/16/165602,PhysRevB.87.054402,1212.6012,1904.02102,PhysRev.165.647,PhysRevB.46.3486,PhysRevB.46.8268,PhysRevB.45.2207,PhysRevB.63.144428} chain with nearest-neighbor (NN) couplings 
\begin{equation}
\begin{aligned}
H &= J_1 \sum_{i=1}^{N/2}   \left(  S_{2i-1}^{x} S_{2i}^{x} 
+  S_{2i-1}^{y} S_{2i}^{y}
+ \Delta_z S_{2i-1}^{z} S_{2i}^{z} \right) \\ 
&+ J_2 \sum_{i=1}^{N/2-1}   \left(  S_{2i}^{x} S_{2i+1}^{x} 
+  S_{2i}^{y} S_{2i+1}^{y}+ \Delta_z S_{2i}^{z} S_{2i+1}^{z} \right)\;,
\label{eq:Hamilton_NN}
\end{aligned}
\end{equation}
where $J_1$ and $J_2$ are the inter- and intra-dimer couplings and
$\Delta_z$ gives their $z$-axis anisotropy. 
$\Delta_z \gg 1$ leads to more Ising-like spins while $\Delta_z \ll 1$ would imply $x$-$y$ anisotropy. 
Topological phases only occur with an AFM~\cite{PhysRevB.87.054402} $J_2 > 0$, see Fig.~\ref{fig:phases_ideal}, and we use $J_2$ as our unit of energy, i.e. $J_2=1$. 
Since we are interested in edge states, we use open boundary conditions (OBC),
so that edge states equivalent to $S=\tfrac{1}{2}$ (see Fig.~\ref{fig:cartoon}(b-d)) imply a four-fold degenerate
ground state~\cite{T_Kennedy_1990,PhysRevB.50.6277,PhysRevB.48.913}. 

These terms are complemented by NNN coupling~\cite{1508.06129,1212.6012,10.7566/JPSJ.85.124712,1904.02102,10.1088/0953-8984/26/27/276002}
\begin{align}
H_{\textrm{NNN}} &=  J_\mathrm{NNN} \sum_{i=1}^{N-2}\left( S_{i}^{x} S_{i+2}^{x} 
+  S_{i}^{y} S_{i+2}^{y} + \Delta_z S_{i}^{z} S_{i+2}^{z} \right).
\label{eq:Hamilton_NNN}
\end{align}
In addition, there can be an uniaxial single-ion anisotropy~\cite{DELGADO201740,inbook_regnault,PhysRevA.93.464,PhysRevB.96.060404,PhysRevB.46.8268,PhysRevB.84.180410,PhysRevB.45.9798} ($D$-anisotropy)
\begin{align}
H_\mathrm{D} = D \sum_{i=1}^{N/2} (S_{2i-1}^{z} + S_{2i}^{z})^2,
\label{eq:D_anis}
\end{align}
where $D$ is the strength of the anisotropy.

A standard method to distinguish between the topologically trivial and
nontrivial phases is the string order parameter~\cite{PhysRevB.87.054402,1904.02102,PhysRevB.46.3486,PhysRevB.46.8268,Bahovadinov_2019,PhysRevB.44.11789} introduced by  Nijs and
Rommelse~\cite{PhysRevB.40.4709} and
Tasaki~\cite{PhysRevLett.66.798}.  However, it is only cleanly defined for
infinite chains, while we are here explicitly interested in finite ones. 
 Moreover, our focus is mostly on the edge states in imperfect systems, where they
arise, and how robust they are. Consequently, we use the (approximate) four-fold degeneracy of the ground
state, which is an important property of the topological state of a chain with OBC, as a criterion. 

We investigate the Hamiltonian by using exact diagonalization (ED) of chains,
whose length is comparable to the number of spins that can be assembled in an
STM~\cite{Mishra2021,GAUYACQ201263,10.1038/s41467-019-10103-5,doi:10.1126/science.1125398,nphys3722,doi:10.1126/science.1214131,doi:10.1021/acs.nanolett.7b02882}. Chains of, e.g., twelve spins are accessible to full numerical
diagonalization. In some cases, we go to longer chains to
assess finite-size effects, which can be strong for short
chains~\cite{T_Kennedy_1990,PhysRevB.87.144409,PhysRevB.50.6277,PhysRevB.48.913}, and then use a band-Lanczos
algorithm~\cite{ruhe_band_lanczos}. In this variant 
of the standard Lanczos approach, several starting vectors are used at the
same time and are mutually orthogonalized. While the plain Lanczos
algorithm cannot reliably resolve (near-)degeneracies~\cite{ruhe_band_lanczos}, we can
thus determine them as long as the number of starting vectors is greater than
the degeneracy.

\section{Results} \label{sec:results}

\begin{figure}
  \includegraphics[width=\columnwidth]{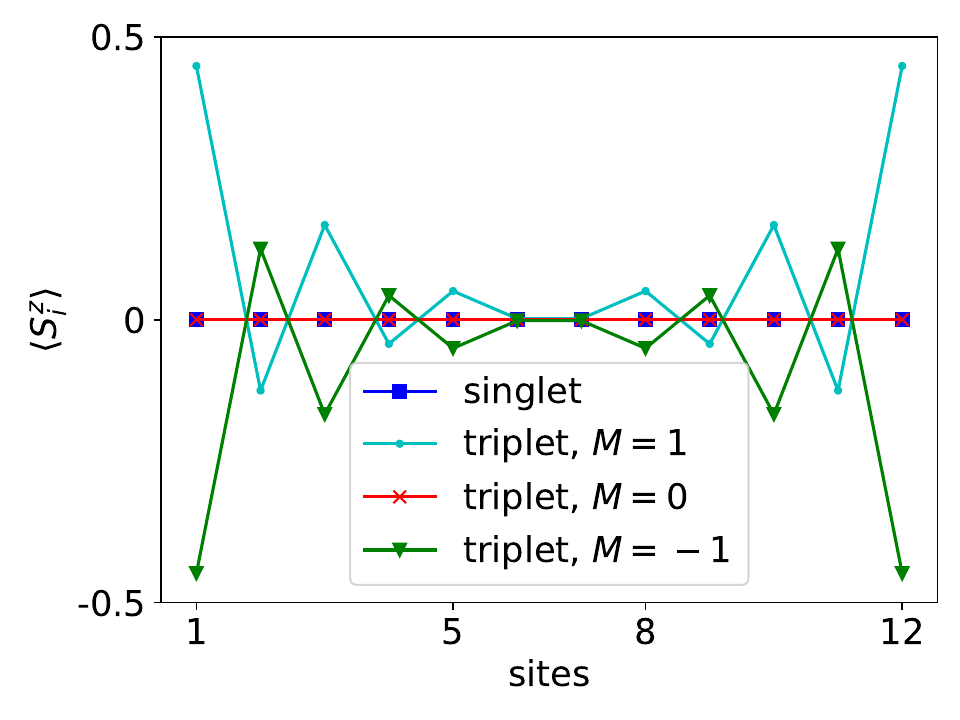}
    \caption{Expectation value $\langle S_i^z \rangle $ of the singlet and triplet states of the edge modes for an isotropic chain consisting of twelve spins,  with $J_1=0.5$ and without NNN couplings. The properties of the four ground states are well recognizable. \label{fig:S_z_exp_val}}
\end{figure}
Fig.~\ref{fig:S_z_exp_val} shows the expectation value $\langle S_i^z \rangle$ of the singlet and triplet states of the edge modes, see Fig.~\ref{fig:cartoon}(d), where $S_i^z $
is the spin operator of the  $i$-th spin in the chain. The triplet states with $M=\pm1$ have, exactly as expected,
$\langle S_i^z \rangle \approx\pm 1/2$ at the edges and
$\langle S_i^z \rangle $ decays exponentially towards the
bulk of the chains. 
The stronger the dimerization (i.e., the smaller $|J_1|/J_2$), the
smaller the difference $||\langle S_i^z \rangle | -1/2|$ at the edges.

\begin{figure}
\includegraphics[width=\columnwidth]{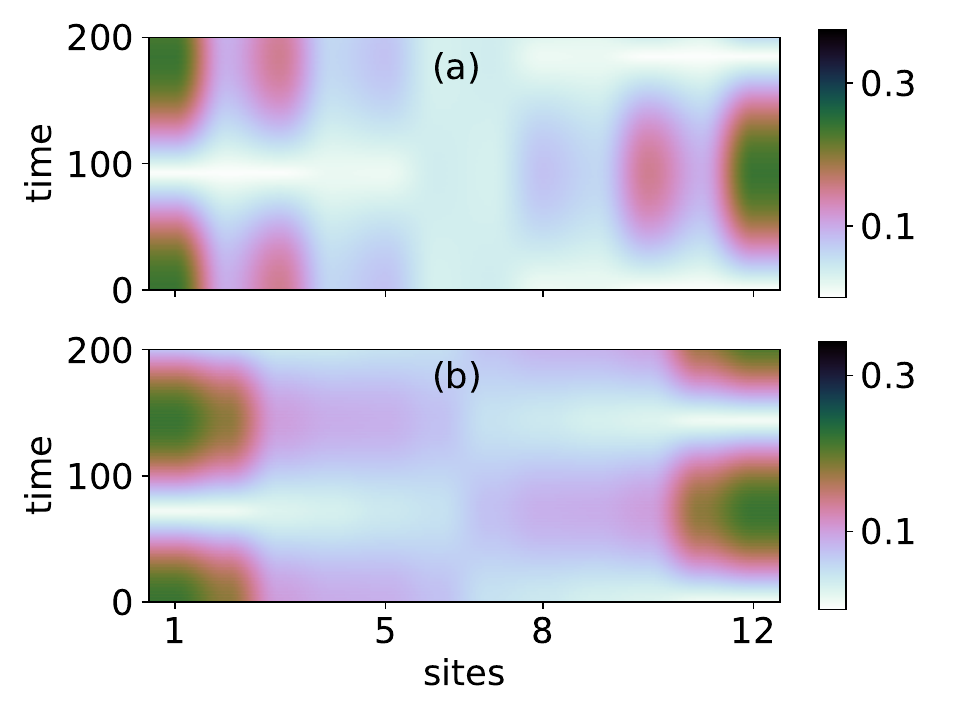}
\caption{Expectation value $|\langle S_i^z \rangle |$ of the time evolution of a state Eq.~\eqref{eq:ini_state} prepared to
  have $\langle S_i^z\rangle=\tfrac{1}{2}$ on the left edge for (a)
  the dimer scenario with $J_1=0.7$ and (b) the Haldane scenario with $J_1=-5$.\label{fig:time} }
\end{figure}

Energy splitting as a criterion can be motivated by the goal of
observing an edge state. For instance, consider the time evolution of
a low-energy state prepared to have spin 'up' at the left edge, which
can be obtained by combining three of the four quasi-degenerate low-energy states, see Fig.~\ref{fig:S_z_exp_val}, as 
\begin{equation}
\begin{aligned}\label{eq:ini_state}
  |i\rangle &= \frac{1}{\sqrt{2}}  \biggl (
  |S=1,M=1\rangle  \\ 
  &+  \frac{1}{\sqrt{2}}(|S=1,M=0\rangle +|S=0,M=0\rangle) \biggr)  \;.
\end{aligned}
\end{equation}
As can be seen in Fig.~\ref{fig:time} at time $t=0$, the $z$-component of the spin on
the right edge then averages to zero.  
Experimentally, one might expect such a state if the temperature is low
enough to bring the chain into its quasi-degenerate low-energy
manifold, but if a magnetic tip polarizes one edge. Time evolution can
then be calculated using ED.

Fig.~\ref{fig:time} shows the time evolution of $\langle S_i^z \rangle$  and compares a Haldane-like scenario (i.e. with a strong FM
$J_1$) to a dimer-scenario. One sees that in both cases, the positive spin remains on the
left edge for relatively long times before tunneling to the right edge. The tunneling time is inversely proportional to the energy difference between the singlet and triplet states,  i.e. directly related to the splitting within the
quasi-degenerate manifold resp. the coupling of the edge modes.

Due to the finite length of the spin chains, there is an energy gap between the singlet and the triplet states,  see Fig.~\ref{fig:S_z_exp_val}.
The singlet state is the ground state if $J_1$ and $J_2$ are AFM. A chain with FM $J_1$ and AFM $J_2$ has a singlet ground state if $N/2$ is even and a triplet ground state if $N/2$ is odd. This can be explained by the  Lieb Schultz Mattis theorem~\cite{PhysRevB.50.13442,10.1063/1.1724276}.

\subsection{Inelastic tunneling current and spin spectral density}\label{sec:current}
Inelastic electron tunneling spectroscopy can be used to measure edge states with an STM~\cite{Mishra2021}.
Excitation from the singlet (triplet) ground state into the nearly
degenerate triplet (singlet) state is found at an energy close to zero if the
edge states are indeed independent of each other. The expected signal
is proportional to the matrix elements~\cite{Mishra2021,PhysRevLett.102.256802,doi:10.1126/science.1146110,PhysRevLett.111.167201} describing such transitions 
\begin{align}\label{eq:S_inel}
S_\mathrm{inel}= \sum_{\alpha,j} |\langle \mathrm{SG}|S_i^\alpha|\mathrm{TP}_j\rangle|^2,
\end{align}
where $\ket{\mathrm{SG}}$ is the singlet state and
$\ket{\mathrm{TP}_j}$ with  $j=1,2,3$ 
are the triplet states formed by the
edge spins, see Fig.~\ref{fig:cartoon}(b-d).  $S_i^\alpha $
is the spin operator for component $\alpha = x,y,z$ of the $i$-th spin in the chain.

\begin{figure}
  \includegraphics[width=\columnwidth]{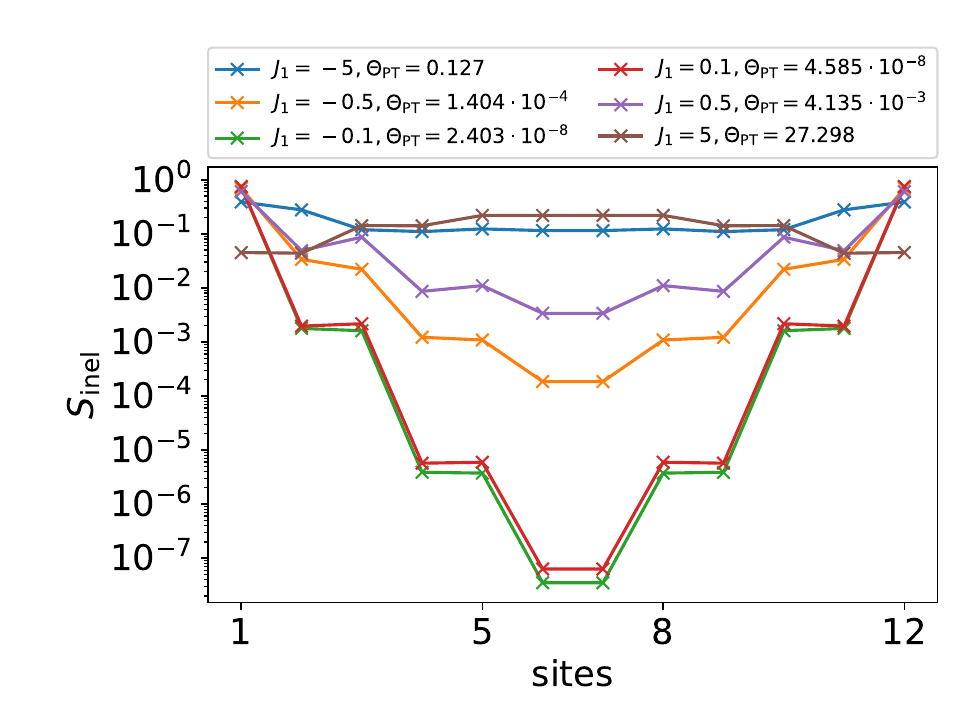}
    \caption{Inelastic signal Eq.~\eqref{eq:S_inel} for isotropic chains with twelve spins
      and without NNN couplings. For comparison between both phase
      transition criteria, there are in the legend the values of
      $\Theta_\mathrm{PT}$, see Eq.~\eqref{eq:def_quasi_deg}.
\label{fig:S_inel_simple_chain}}
\end{figure}

For a spin-1 Haldane chain, the signal is indeed found to be stronger
on the first and last spin~\cite{PhysRevLett.111.167201}, by the ratio of
about four, so that it can be used to detect edge
states. Fig.~\ref{fig:S_inel_simple_chain} shows the matrix-element
weight Eq.~\eqref{eq:S_inel} for a twelve-site chain and varying $J_1$. In
the topological regime $J_1 < J_2$, weight is consistently concentrated
on the edges, while it is found in the bulk for the trivial $J_1 =
5\;J_2$. One also sees that the localization of the edges becomes more
pronounced in the dimer regime with $|J_1| < J_2$ than it is in the
Haldane regime with $J_1 < -1$, see Fig.~\ref{fig:phases_ideal}. This suggests that dimerized chains
might be a good place to investigate edge states.

\begin{figure}
  \includegraphics[width=\columnwidth]{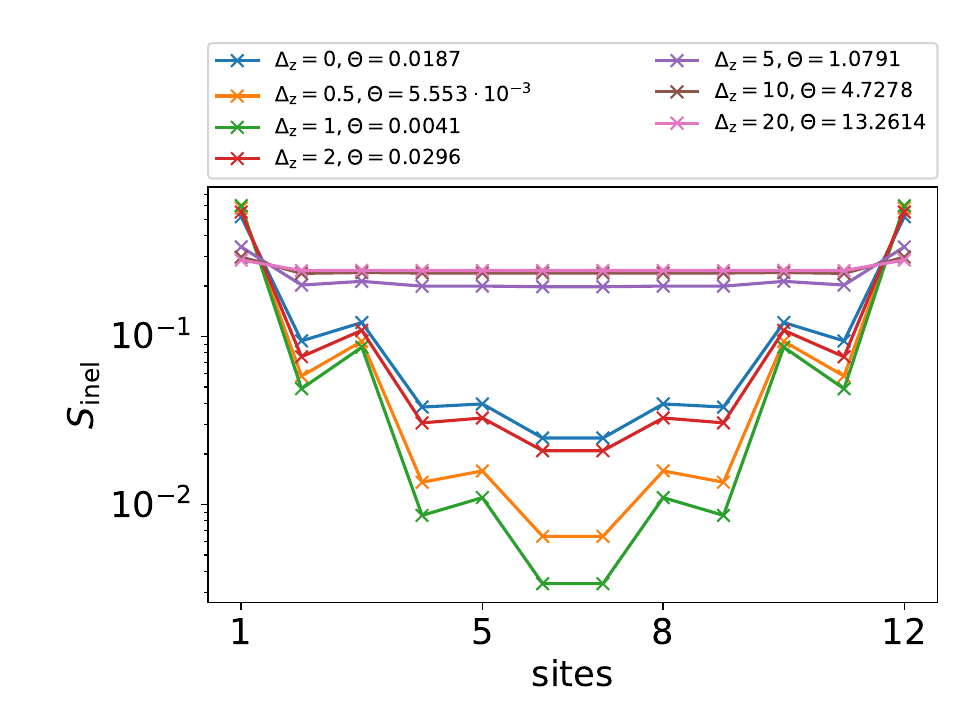}
    \caption{Inelastic signal Eq.~\eqref{eq:S_inel} for an anisotropic chain with twelve spins,  without NNN couplings and $D$-anisotropy. The coupling constant is $J_1=0.5$. For comparison between both phase transition criteria, there are in the legend the values of  $\Theta_\mathrm{PT}$, see Eq.~\eqref{eq:def_quasi_deg}.
\label{fig:S_inel_Delta_z}}
\end{figure}

Fig.~\ref{fig:S_inel_Delta_z} shows the impact of a $z$-axis anisotropy on
the inelastic signal Eq.~\eqref{eq:S_inel} for a dimerized chain. The signal
continues to be located at the edges also for strong anisotropies $\Delta_z$ (see
Eq.~\eqref{eq:Hamilton_NN}). It is possible to drive the excitation from the
edge into the bulk, but very strong anisotropies (e.g. $\Delta_z =
10$) are needed to achieve a signal (almost) uniformly
distributed over the chain.

However, even when transitions between the four lowest states are localized at the
edges, the site-dependent inelastic signal Eq.~\eqref{eq:S_inel} can only
be useful if any bulk excitations are at appreciably higher energies. This
translates into the requirement that the four low-energy states are 
separated from the rest of the spectrum by a gap. Information on edge states and energy gaps is combined in Green's
functions, i.e., in dynamic spectra of the form
\begin{align}
S_i(\omega) = \sum_{m,\alpha} |\langle m|S_i^{\alpha}|0\rangle|^2 \delta(\omega-(E_m-E_0))\;.
\label{eq:dynamic_spectra}
\end{align}
Operator $S_i^{\alpha}$  denotes as above spin
component $\alpha=x,y,z$ at site $i$ and $|0\rangle$ is the ground state with
energy $E_0$. In contrast to Eq.~\eqref{eq:S_inel} above, the sum runs over
all excited states 
$|m\rangle$ with energy $E_m$, not just the quasi-degenerate
triplet. We use the Lanczos algorithm to numerically obtain
spectra.

\begin{figure}
\includegraphics[width=\columnwidth]{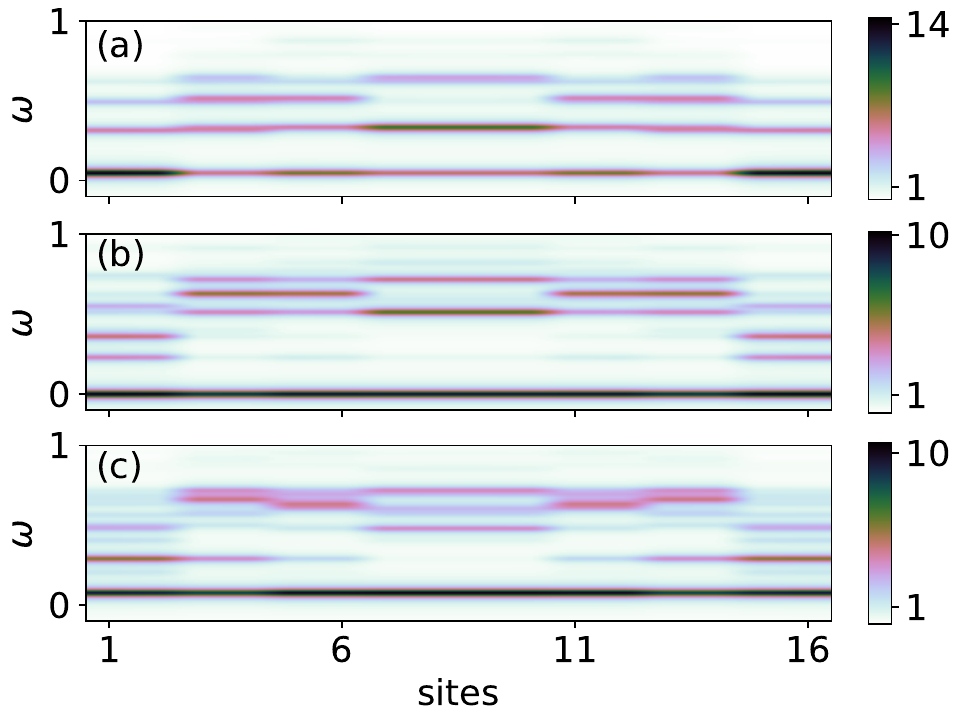}
\caption{Site-dependent spin-excitation spectra  Eq.~\eqref{eq:dynamic_spectra} for  $J_1=-5$ (Haldane
  scenario). (a) is the isotropic chain, i.e. $\Delta_z=1$, (b)
  $\Delta_z=1.01$, and (c) $\Delta_z=0.99$. $J_{\textrm{NNN}}=0$ and $D=0$ in
  all cases. \label{fig:spectra_haldane}}
\end{figure}

\begin{figure}
\includegraphics[width=\columnwidth]{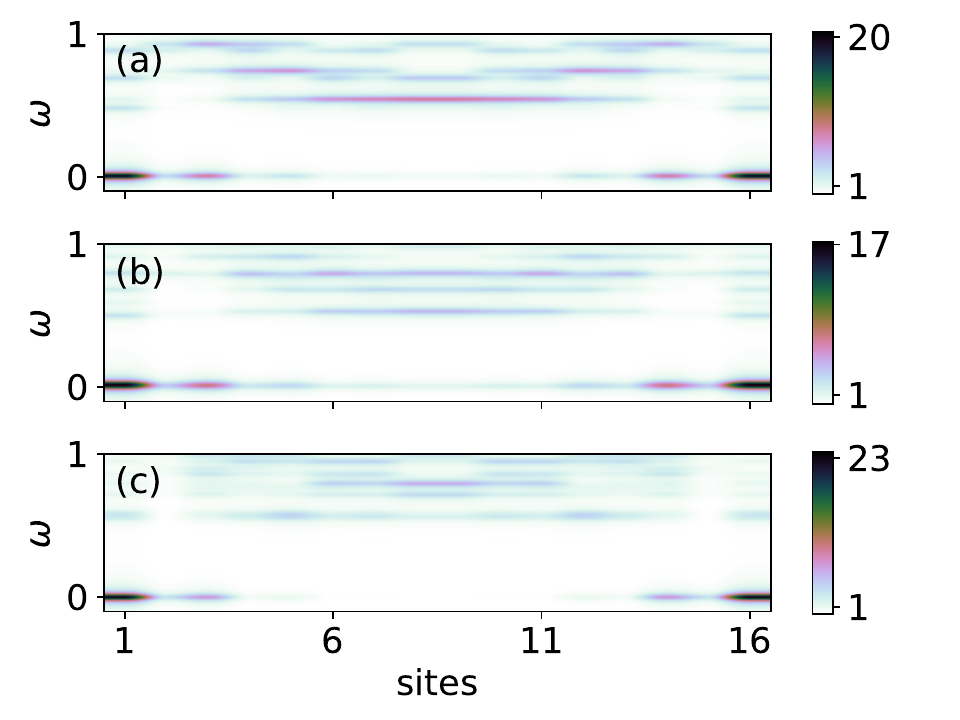}
\caption{Site-dependent spin-excitation spectra Eq.~\eqref{eq:dynamic_spectra} for $J_1=0.7$
  (dimer scenario). The remaining parameters are (a) $J_{\textrm{NNN}}=0.0$,
  $\Delta_z=1$, (b) $J_{\textrm{NNN}}=0.0$,
  $\Delta_z=1.5$, and  (c) $\Delta_z=1.5$, $J_{\textrm{NNN}}=0.2$. $D=0$ in all cases. \label{fig:spectra_dimer}}
\end{figure}

Spectra shown in Fig.~\ref{fig:spectra_haldane} for a Haldane-like
chain again show the low-energy
excitations located almost exclusively at the end sites, with
only a little weight leaking into the chain. Higher-energy excitations
into states beyond the quasi-degenerate ground state manifold are also
seen within the chain, their energy corresponds to the gap between the
fourth and fifth eigenstates of the chain (see Fig.~\ref{fig:closed_lowest_energies_2}). As found
previously~\cite{PhysRevLett.111.167201}, even rather small $z$-axis
anisotropy of $\approx 1\;\%$ is enough to allow a substantial part of
the edge state weight to leak into the bulk, see
Fig.~\ref{fig:spectra_haldane}(b,c).

The higher-energy
states that come down in energy at the same time are, however, also
found rather close to the edges. This implies that the identification of
edge states using the inelastic signal can only be a first step: The
site-dependent low-energy weight can still be enhanced near the edges,
even though the object at the edge consists of several excitations and
is no longer similar to a spin $1/2$.

Fig.~\ref{fig:spectra_haldane}(b-c) and Fig.~\ref{fig:spectra_dimer}(b-c) compare the
impact of $z$-anisotropy $\Delta_z$ (see Eq. \eqref{eq:Hamilton_NN}) on the dimer and Haldane
scenarios. The low-energy edge states move into the chain already for
small deviations from $\Delta_z=1$ in the Haldane scenario, see Fig.~\ref{fig:spectra_haldane}(b-c). In contrast, they remain clearly localized in the dimer cases Fig.~\ref{fig:spectra_dimer}(b-c), just as in the case of the isotropic dimer scenario Fig.~\ref{fig:spectra_dimer}(a).

\subsection{Energy gaps as a criterion for edge states}\label{subsec:gaps}

\begin{figure}
\includegraphics[width=\columnwidth]{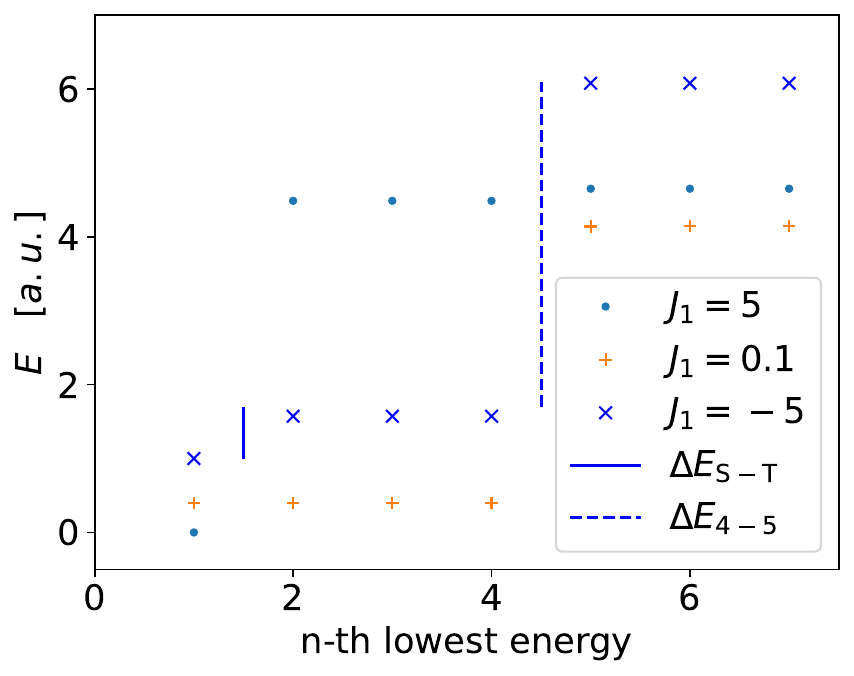}
\caption{Lowest few energies depending on $J_1$ for an isotropic spin chain with twelve spins and  $J_{\textrm{NNN}}=0$. 
	There are two important energy gaps.
 	First, $\Delta E_\mathrm{S-T}$, the energy gap between the singlet and the triplet states,
	 and second  $\Delta E_\mathrm{4-5}$, the energy gap between the fourth and the fifth lowest energy.
	The gap $\Delta E_\mathrm{S-T}$ is tiny in the case of a strong dimer formation $J_1=0.1$. 
	For a weak dimer formation and in the Haldane phase, e.g. $J_1=-5$, this energy gap is increasing, but there remains a quasi four-fold degenerate ground state.
	In the topologically trivial phase, e.g. $J_1=5$, there is a nondegenerate ground state.}
	\label{fig:closed_lowest_energies_2} 
\end{figure}

The information on energy gaps discussed in the figures above can be summarized
into a parameter given by the ratio of gaps in the eigenvalue spectrum of the
Hamiltonian. Left-over spins $S=\tfrac{1}{2}$ at the edges, see
Fig.~\ref{fig:cartoon}(b-d), which are decoupled for infinite chains, will
become coupled in realistic short chains, which  splits the
four-fold degenerate ground state into a singlet and a
triplet~\cite{Mishra2021,T_Kennedy_1990,PhysRevB.50.6277,PhysRevB.48.913}. To consider
the edge states as still 'approximately decoupled', the singlet-triplet gap 
should be smaller than the  gap $\Delta E_{4-5}$ separating the
fourth from the fifth state, see Fig.~\ref{fig:closed_lowest_energies_2}. 

Fig.~\ref{fig:closed_lowest_energies_2} illustrates the
concept, see Eq.~\eqref{eq:def_quasi_deg}, to distinguish between
topological and topologically trivial phases.  
One see example spectra for two cases with an approximate ground-state
degeneracy ($J_{1}=-5$ and $J_{1}=0.1$) and
one without edge states ($J_{1}=5$). 
We  evaluate 
\begin{align}
\Theta_\mathrm{PT}  \coloneqq
\left\{
\begin{matrix}
\max\limits_{\substack{i=1,2,3 }} \frac{\Delta E_i}{\Delta E_{4-5}}&
 \hspace{0.5cm} \mathrm{if\ } D_\mathrm{GS} \leq 4  \\
\frac{\Delta E_\mathrm{GS}}{\Delta E_{4-5}}&  \hspace{0.5cm} \mathrm{otherwise,} \\
\end{matrix}
\right. 
\label{eq:def_quasi_deg}
\end{align}
where $ D_\mathrm{GS}$ is the (quasi-) degeneracy of the ground state
manifold and  $\Delta E_\mathrm{GS}$ is the energy gap above the ground
state. More generally, the singlet-triplet gap will later be replaced by the
largest gap within the lowest four states. 

We then define transitions where the ratio  $\Theta_\mathrm{PT}$ crosses
a certain value $\Theta_\mathrm{PT}^\mathrm{tran}$. This serves as an
indicator where edge states might be expected to be sufficiently protected
to be observable. While the exact value of the threshold is certainly open to
debate, our primary outcomes do not qualitatively change when
$\Theta_\mathrm{PT}^\mathrm{tran}$ is varied. 
Here, we  use $\Theta_\mathrm{PT}^\mathrm{tran}=0.5$.
To illustrate the use of the parameter, $\Theta_\mathrm{PT}$ values (see Eq.~\eqref{eq:def_quasi_deg}) for
each chain are given in 
Figs.~\ref{fig:S_inel_simple_chain} and~\ref{fig:S_inel_Delta_z}. $\Theta_\mathrm{PT}\ll
\Theta_\mathrm{PT}^\mathrm{tran}$ consistently holds for spin chains where the
inelastic signal is localized at the edges. Spin chains with an inelastic signal concentrated at the edges always have a
quasi-degenerate ground state, while no such degeneracy exists for spin chains with a large inelastic signal
in the bulk.

\subsection{Gaps for the isotropic alternating chain}\label{subsec:simple_chain}

\begin{figure}
\includegraphics[width=\columnwidth]{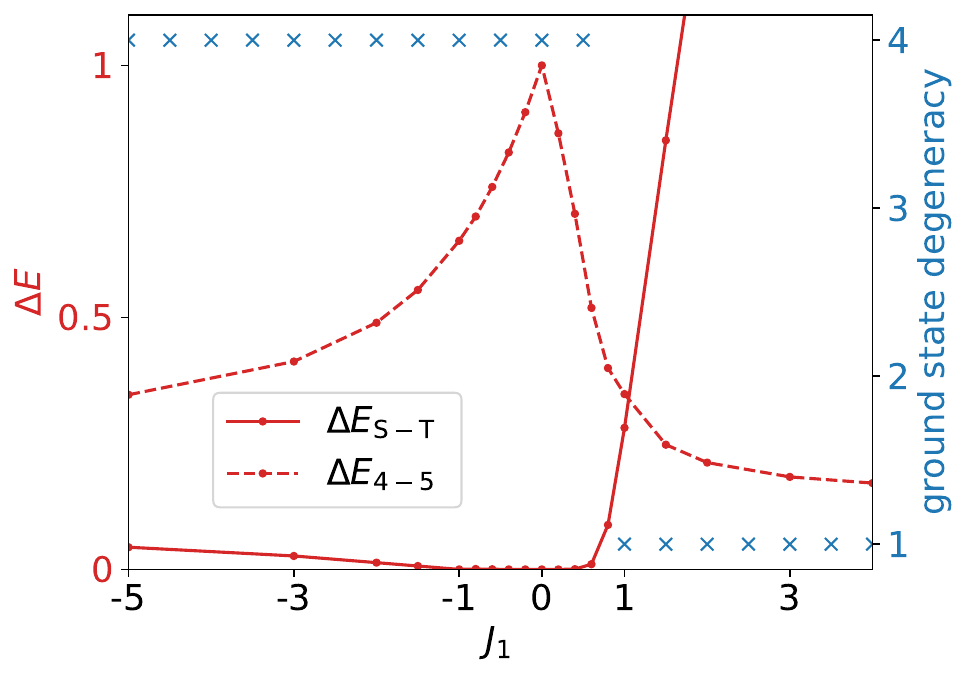}
\caption{Lowest few energies depending on $J_1$ for an isotropic spin chain with twelve spins and  $J_{\textrm{NNN}}=0$. 
	There are shown the energy gaps $\Delta E_\mathrm{S-T}$ and $\Delta E_\mathrm{4-5}$,  as introduced in Fig.~\ref{fig:closed_lowest_energies_2}.
	The gap $\Delta E_\mathrm{S-T}$ is tiny in the case of a strong dimer formation $|J_1| \ll 1 $. 
	For a weak dimer formation and in the Haldane phase this energy gap is increasing, but there remains a quasi four-fold degenerate ground state.
	In the topologically trivial phase, i.e. $J_1\geq1$, there is a nondegenerate ground state.
	The four-fold degeneracy resolves at the topological phase transition
        at $J_1=1$. }
	\label{fig:closed_lowest_energies_1}
\end{figure}

\begin{figure}
\includegraphics[width=\columnwidth]{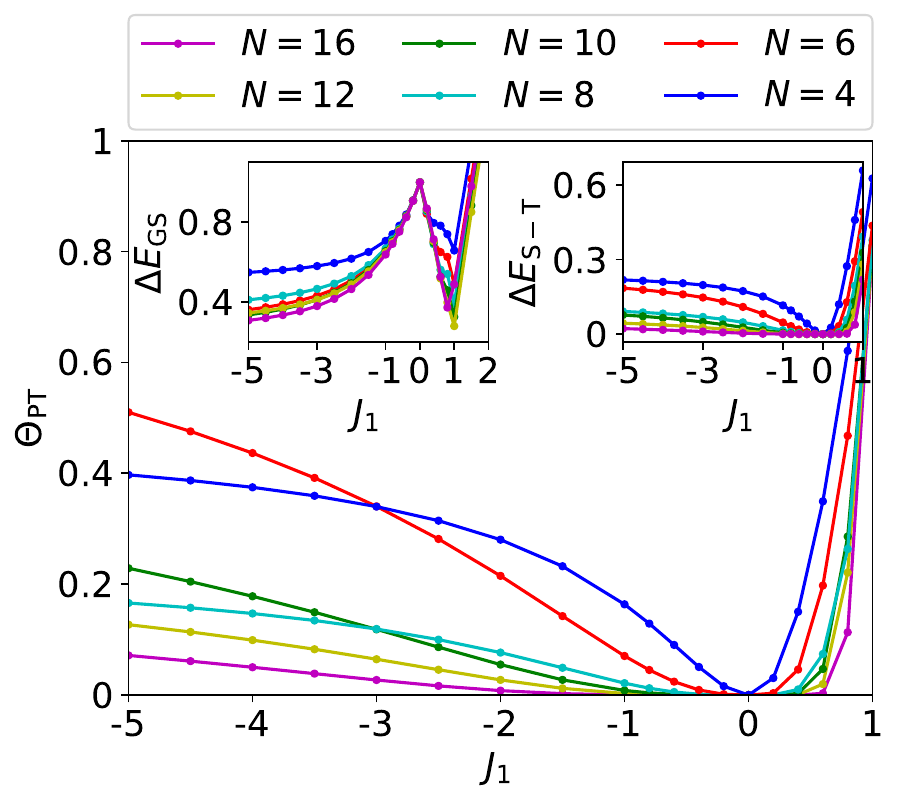}
\caption{Edge state interaction depending on chain length. The results belong to an isotropic spin chain without NNN coupling.
  	The left inset shows the energy gap above the ground state $\Delta E_\mathrm{GS}$. This gap has a minimum $J_1=1$. 
	The right inset shows $\Delta E_\mathrm{S-T}$ (see Fig.~\ref{fig:closed_lowest_energies_2}).
	The ratio, see Eq.~\eqref{eq:def_quasi_deg}, between the energy gap inside the quasi-degenerate ground state and the energy gap above the ground state is shown in the large figure.
	\label{fig:closed_size_gaps}}
\end{figure}

Fig.~\ref{fig:closed_lowest_energies_1} shows the two crucial energy gaps,
as introduced in Fig.~\ref{fig:closed_lowest_energies_2}, depending on
$J_1$. For negative (i.e. FM) value of $J_{1}$, the  $\Delta E_\mathrm{S-T}$
is always much smaller than $\Delta E_{4-5}$, so that the system remains in
the topological (Haldane) phase. However, $\Delta E_\mathrm{S-T}$ grows and
$\Delta E_{4-5}$ shrinks, reflected in the ratio
Eq.~\eqref{eq:def_quasi_deg} and making the system less suitable for observation of
the edge states. For positive $J_{1}$, in contrast, we find a sudden transition
from quite small $\Theta_\mathrm{PT}$ to $\Theta_\mathrm{PT}>1$ at the
topological phase transition, where the quasi four-fold ground-state degeneracy is lost.  

Since finite spin chains are analyzed, finite-size effects are a concern.
Fig.~\ref{fig:closed_size_gaps} shows the energy gaps discussed in
Fig.~\ref{fig:closed_lowest_energies_2} as well as their ratio
$\Theta_\mathrm{PT}$, depending on the chain length $N$. While the gap $\Delta
E_{4-5}$ remains finite for longer chains, $\Delta E_\mathrm{S-T}$ is strongly
suppressed within the topological regime once chains become longer than about eight
sites. This suggests that the edges can be considered to be approximately
decoupled even on rather short chains. The ratio given
in the main panel of Fig.~\ref{fig:closed_size_gaps} gives accordingly a
consistent picture except for the very shortest chains.

\subsection{Next-nearest-neighbor coupling}\label{subsec:J_NNN}

Since the gap ratio Eq.~\eqref{eq:def_quasi_deg} summarizes information on the
relevant energy scales conveniently and also works for finite chains, we use
it to delineate regimes where one can hope to observe edge states even in
imperfect systems. We first look at the impact of NNN coupling, see
Eq.~\eqref{eq:Hamilton_NNN}, as analogous hopping is known to destroy
ground-state degeneracy for non-interacting bosons. 

\begin{figure}
  \includegraphics[width=\columnwidth]{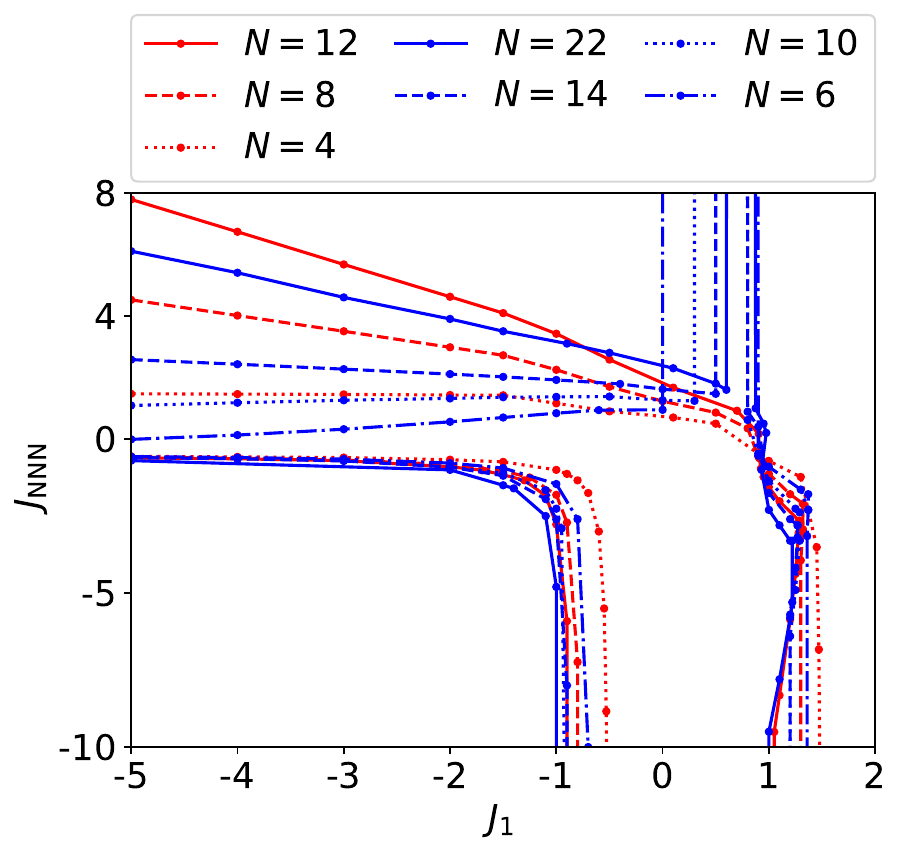}
  \caption{NNN coupling influence depending on chain length. 
    In the topological dimer phase, much stronger FM NNN couplings are possible than in the Haldane phase.
    The finite size effects for AFM NNN coupling show a fundamental difference between $N/2$ is even and $N/2$ is odd. }
  \label{fig:closed_J_NNN}
\end{figure}

For FM NNN coupling, substantial differences between 'Haldane' and 'dimer'
regimes arise, see Fig.~\ref{fig:closed_J_NNN}. In the former, where stronger $|J_1| > |J_2|$ stabilizes the
quasi-spin-1 constituents (see Fig.~\ref{fig:cartoon}(a-b)), even rather weak $J_\mathrm{NNN}< 0$ is enough to
drive a topological phase transition to a (trivial) FM state.
In the topological dimer phase, (i.e. where AFM $J_2$ is stronger than $|J_1|$), there
is no such transition. Thus the topological dimer phase is significantly more
robust than the Haldane phase against FM NNN coupling. 

For AFM NNN coupling, finite-size effects are substantial and show
pronounced even/odd differences, i.e. between even and odd $N/2$, especially
in the Haldane phase.  Depending on chain length, either small values of
$J_{\mathrm{NNN}}> 0$ couple edge spins, or even quite large ones leave them
uncoupled. For long chains and strong FM $J_1$, AFM $J_{\mathrm{NNN}}> 0$ can,
however, be allowed to become quite large without affecting edge state
degeneracy. This fits with earlier observations that NNN coupling can be used
to tune edge-state coupling in Haldane chains~\cite{PhysRevB.96.060409,PhysRevB.97.174434}.

\subsection{Spin anisotropy $\Delta_z$ of couplings and uniaxial
  single-ion--like anisotropy $D$}\label{subsec:z_anis}

\begin{figure}
\includegraphics[width=\columnwidth]{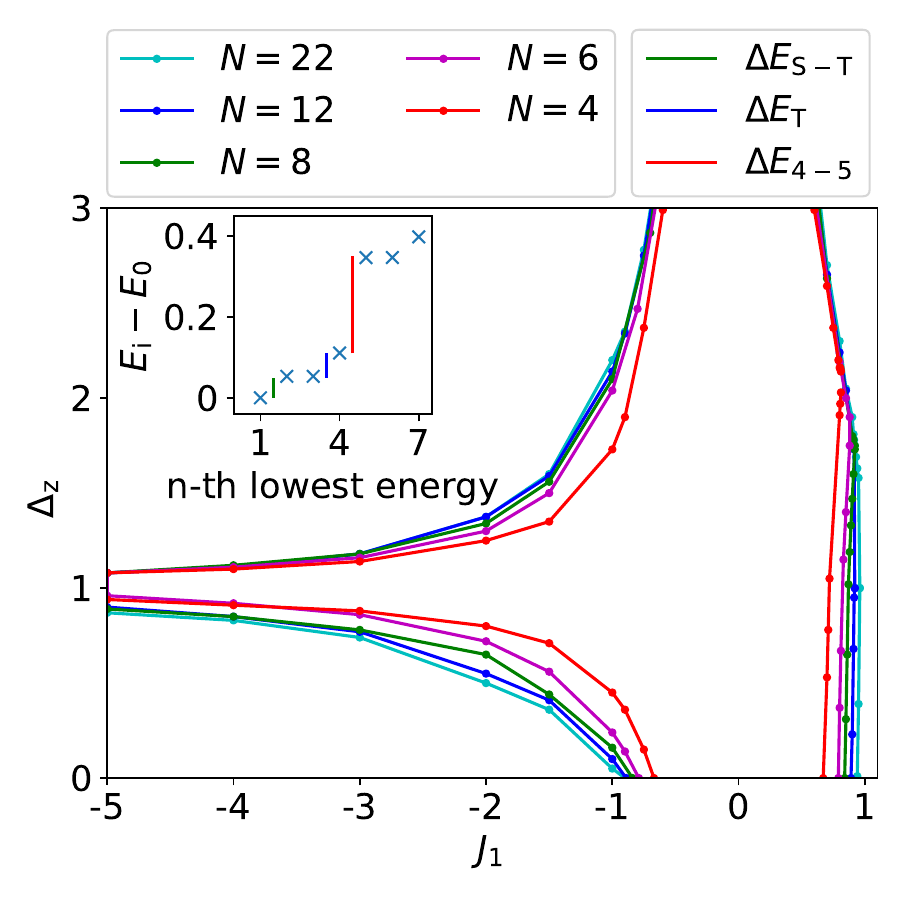}
\caption{$z$-anisotropy influence depending on the chain length for chains without NNN coupling and $D$-anisotropy. 
	The topological dimer phase is more robust than the Haldane phase.
	The finite size effects are small. 
	The inset shows the low energy spectrum, which is shifted with the ground state energy $E_0$.}
	\label{fig:closed_z_anis}
\end{figure}

Next, we will address spin anisotropies that select a specific axis. The first 
purpose is to investigate the connection to the hard-core bosons of
Ref.~\cite{SSH_Rydberg_19}, which correspond to $\Delta_z=0$
in our spin model. The second reason is that such anisotropies have to
be expected for spins on surfaces~\cite{GAUYACQ201263,doi:10.1126/science.1125398,DELGADO201740,doi:10.1126/science.1082857,doi:10.1126/science.1214131,doi:10.1126/science.1146110,PhysRevLett.111.127203,PhysRevLett.102.256802,doi:10.1021/acs.nanolett.7b02882} and have been shown to strongly
affect edge states of Haldane chains~\cite{PhysRevLett.111.167201}.

We model two sources of anisotropy: (i) coupling anisotropy $\Delta_z$ affecting
all spin-spin couplings, see Eq.~(\ref{eq:Hamilton_NN}), and (ii)
'single-ion' anisotropy $D$ of Eq.~(\ref{eq:D_anis}) 
affecting only the $J_1$ bonds. Turning these bonds FM moves our model
into the 'Haldane-like' phase and $D<0$ ($D>0$) then favors states
with $S_z=\pm 1$ ($S_z= 0$)  of a local triplet, as a single-ion
anisotropy~\cite{PhysRevLett.111.167201,PhysRevB.50.6277} would.

Fig.~\ref{fig:closed_z_anis} shows the influence of a $z$-anisotropy on chains without NNN coupling or $D$-anisotropy.
Due to $z$-anisotropy $\Delta_z$, see the Hamiltonian Eq.~(\ref{eq:Hamilton_NN}), the triplet splits in a singlet and a doublet,
as shown in the inset of Fig.~\ref{fig:closed_z_anis}. 
There is thus an additional energy gap $\Delta E_\mathrm{T}$ within
the  quasi-degenerate ground state manifold, in addition to the energy
gaps of Fig.~\ref{fig:closed_lowest_energies_2}. 
The phase diagram in the main panel has only a relatively minor finite size
effects, especially in comparison to the large finite size effects for
NNN coupling, as discussed in Fig.~\ref{fig:closed_J_NNN}. More importantly, edge state degeneracy can be found for a large part of the phase diagram.

For $\Delta_z=0$, which corresponds to hard-core bosons, the
topological regime with detectable edge state degeneracy is mostly
confined to $|J_1| \lesssim J_2$. For $\Delta_z> 0$ which has to be
expected for spin implementations, edge states are first stabilized
up to the isotropic regime $\Delta_z=1$. Beyond this point, i.e., for
Ising-like anisotropy, the stability regime of edge state degeneracy
shrinks again but remains sizeable, see Fig.~\ref{fig:closed_z_anis}.

\begin{figure}
\includegraphics[width=\columnwidth]{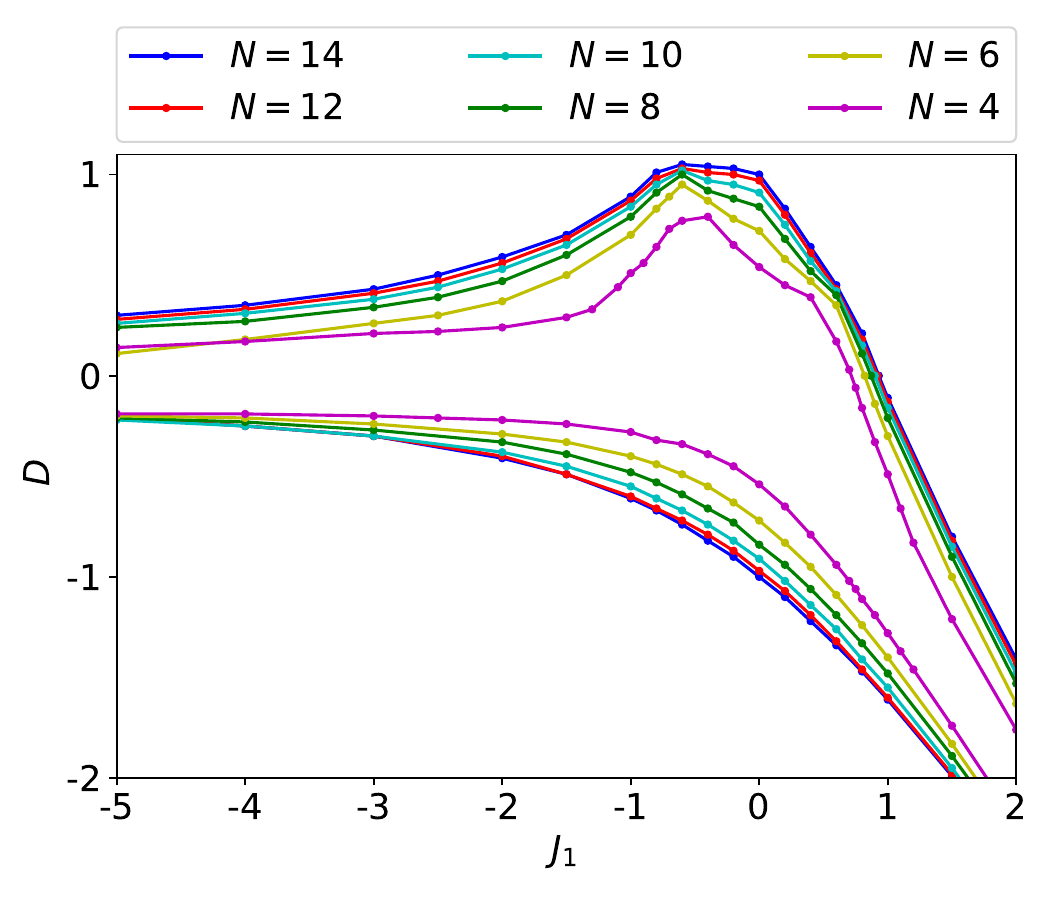}
\caption{$D$-anisotropy influence depending on the chain length for chains without NNN coupling and $z$-anisotropy. 
	The topological dimer phase is more robust than the Haldane phase.
	The finite size effects are small.}
	\label{fig:closed_D_anis}
\end{figure}

Fig.~\ref{fig:closed_D_anis} shows the influence of 
single-ion-like $D$-anisotropy, see Eq.~\eqref{eq:D_anis}. Again,
finite-size effects are moderate, although the topological phase
increases with chain length. Similar to $\Delta_z$ (see Fig.~\ref{fig:closed_z_anis}), edge state degeneracy
is remarkably robust. Again, this robustness is most pronounced for
the dimer regime, while edge states in the Haldane-like regime at
$J_1 < -1$ (see Fig.~\ref{fig:phases_ideal}) are more easily destroyed.

The most striking and relevant result is that the
topological dimer phase is remarkably robust against both types of 
$z$-anisotropy. Similar to FM NNN coupling, this robustness is
enhanced even for FM $J_1$. Edge state degeneracy is here much better
protected than in the Haldane--like spin-1 regime at large $|J_1| \gg 1$, see Fig.~\ref{fig:closed_z_anis}. Robustness extends to
$\Delta_z < 1$ ($x$-$y$ symmetry) and $\Delta_z > 1$
(Ising symmetry). The latter is particularly encouraging because
spins on surfaces usually have a $z$-anisotropy~\cite{doi:10.1126/science.1125398,DELGADO201740,doi:10.1126/science.1082857,doi:10.1126/science.1214131,doi:10.1126/science.1146110,PhysRevLett.111.127203,PhysRevLett.102.256802,doi:10.1021/acs.nanolett.7b02882}.

\subsection{$z$-anisotropy and NNN coupling}\label{sec:z_anis_NNN}

\begin{figure}
          \includegraphics[width=0.5\textwidth]{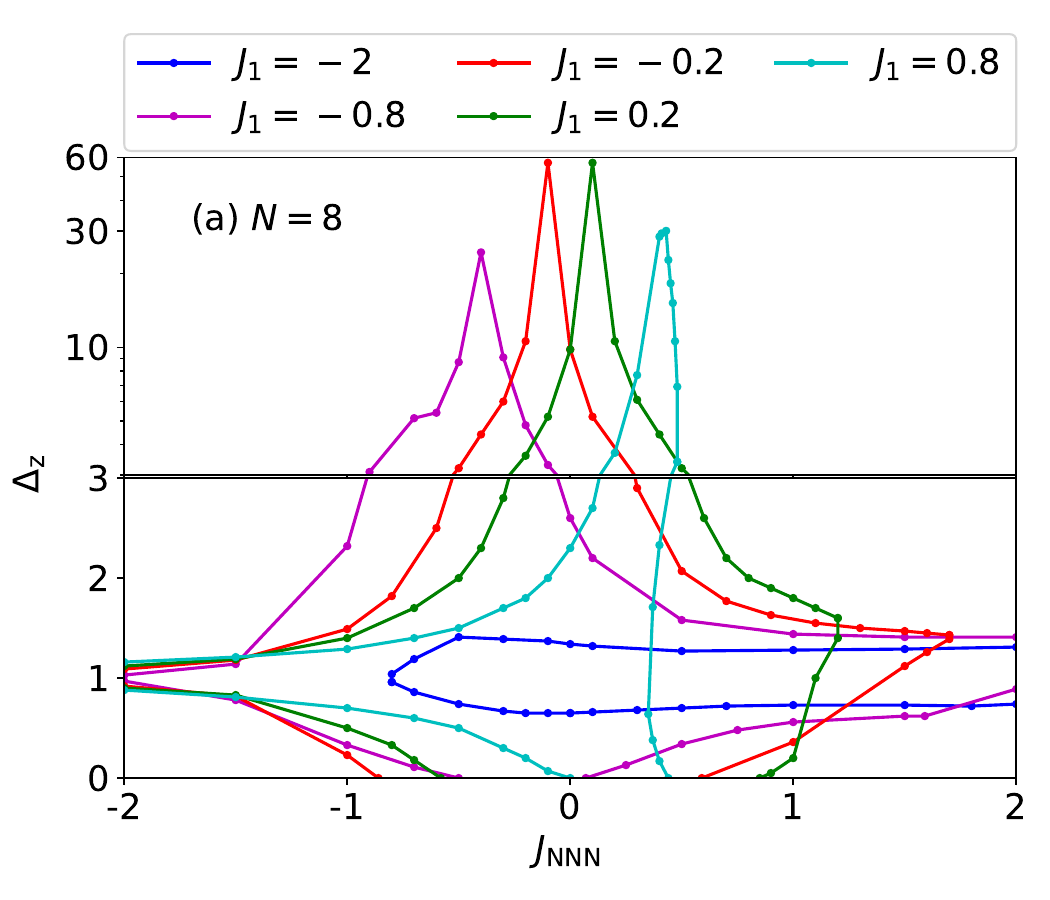}
              \includegraphics[width=0.5\textwidth]{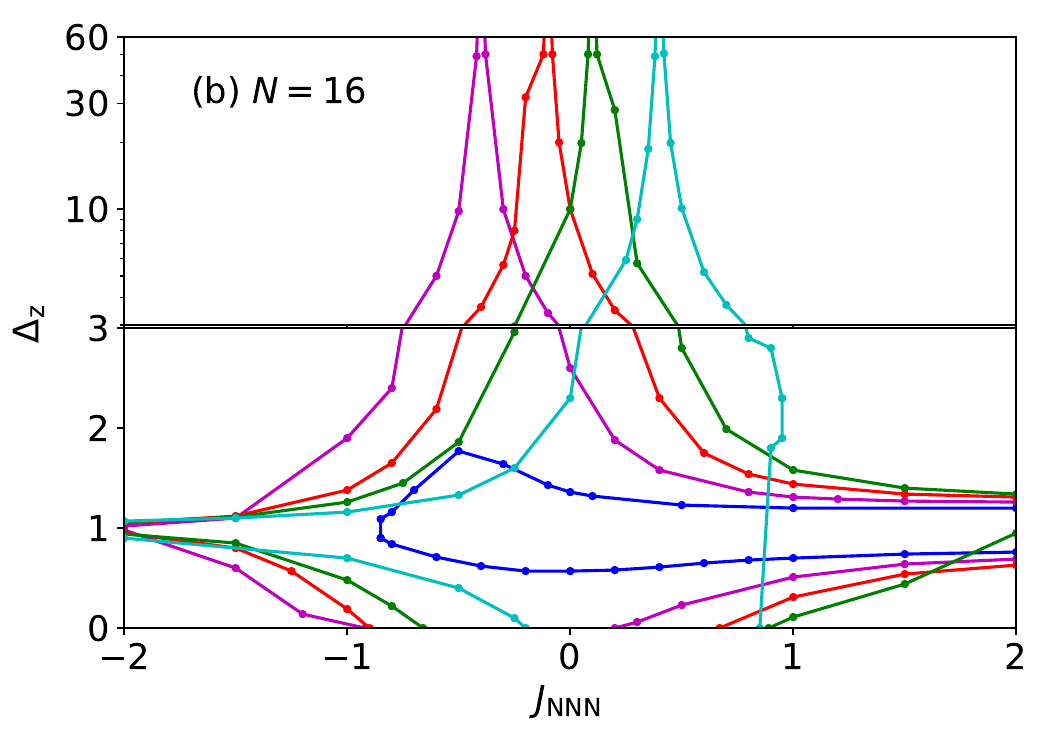}
    \caption{ Influence of the combination of a $z$-anisotropy with NNN coupling for a spin chain with eight spins in (a) and sixteen spins in (b).}
    \label{fig:closed_J_NNN_z_anis}
\end{figure}

Finally, we consider the interplay of NNN coupling and $z$-axis spin
anisotropy. As a result of a 'single-ion-like' anisotropy
affecting only $J_1$ bonds are similar to $\Delta_z$ affecting all
bonds, see above in Sec.~\ref{subsec:z_anis}, we focus here on $\Delta_z$ anisotropy, see
Eq.~(\ref{eq:Hamilton_NN}). 

The stability regions of quasi-degenerate edge states are shown for
various values of $J_1$ and eight-site chains in
Fig.~\ref{fig:closed_J_NNN_z_anis}(a) and with sixteen spins in Fig.~\ref{fig:closed_J_NNN_z_anis}(b). 
Data inherit here the pronounced finite-size
effects observed for NNN coupling in Sec.~\ref{subsec:J_NNN}, as well
as the even/odd effects with odd or even $N/2$.
For visibility, the $\Delta_z$ axis is split into
small and large regimes in
Fig.~\ref{fig:closed_J_NNN_z_anis}. 

For the Haldane regime with $J_1 = -2$, only relatively small
$\Delta_z$ and FM $J_\mathrm{NNN}<0$ are allowed, while a larger range of AFM $J_\mathrm{NNN}>0$ are
acceptable, both in agreement with results discussed earlier (see Fig.~\ref{fig:closed_J_NNN} and Fig.~\ref{fig:closed_z_anis}). In the
dimer regime $J_1 = -0.8$, stability w.r.t. $\Delta_z$ increases
immediately and markedly. $J_\mathrm{NNN}$ must now not become quite as
large, but would still need to be unrealistically large $J_\mathrm{NNN}> J_1$
to destroy quasi-degeneracy. Moreover, $J_\mathrm{NNN} < 0$ is now also
possible, which again illustrates that edge states are far better
protected in the dimer regime.

Even for very large values of $\Delta_z\gg 1$, the protected edge
states can be found for some (albeit narrow) range of
$J_\mathrm{NNN}\approx J_1/2$. However, this only applies to the dimer
regime, but not to Haldane-like chains. This may be related to the
frustration, caused by competing NN and NNN couplings. 
When comparing the robustness between chains with and without frustration, 
it can be seen that frustrated chains tend to have more robust
edge state degeneracy, so that a larger $|\Delta_z-1|$ is possible.

\section{Discussion and Conclusions}\label{sec:discuss}

We have investigated how well edge states of topologically nontrivial
one-dimensional spin chains are protected in finite chains and for
deviations from the ideal model Hamiltonians. To do so, we compared
the energy gap(s) within the four lowest eigenstates, which is due to
coupling of the edge states and vanishes for the ideal case, to the
'topological' energy gap separating these four states from the rest of
the spectrum.

We focus our investigations explicitly on finite chains, as we aim
to identify regimes where topological degeneracy is visible on chains
short enough to be assembled in an
STM. While there are significant
finite-size effects when AFM NNN couplings are involved, even relatively
short chains support parameter regimes with reasonably protected
edge state degeneracy, i.e., where the splitting within the four
lowest-energy states is smaller than their separation from the rest of
the spectrum. 

In comparison to the strong finite size effects for AFM NNN couplings, for FM
NNN couplings the finite size effects are small. A phase diagram for FM NNN
coupling for the 'Haldane-like' scenario of alternating AFM and FM couplings
was obtained with DMRG calculations~\cite{1212.6012} for large chains and then
extrapolated to infinite chains. The infinite chain phase diagram is broadly
consistent with our finite chain phase diagram, which further supports the
outcome of small finite size effects for FM NNN couplings. 

Our main conclusion is that the topological degeneracy is generally more robust
for chains with alternating AFM couplings, i.e. a spin variant of the
SSH model, than for models mimicking Haldane chains, i.e., for
alternating FM and AFM couplings.  Higher robustness means here
that, e.g., larger NNN couplings or stronger $z$-axis anisotropies can
be added to the ideal chain without significantly coupling the edge
states and thus destroying the topological phase.

We have also discussed $z$-axis anisotropy for both alternating coupling
constants because we wanted to discuss spin chains on surfaces and the
direction perpendicular to the surface is clearly
special. A slightly different variant of $z$-axis anisotropy, namely one
applying only to $J_2$ and not to $J_1$, was investigated with disorder operators based on a
cluster expansion~\cite{PhysRevB.48.9555} and with the string order parameter
for periodic boundary conditions~\cite{PhysRevB.46.3486}. 
Although the z-axis anisotropy is distinct, the clear difference in robustness
between the Haldane limit and the dimer regime is in good agreement with our
results based on energy gaps for OBC.  

Concerning the uniaxial single-ion anisotropy $D$ for the $z$-axis, these topological regimes agree reasonably well with the topological
phase diagram for infinite
chains~\cite{PhysRevA.93.464,PhysRevB.46.8268}. Such phase diagrams were obtained for the
'Haldane-like' scenario of alternating AFM and FM couplings. Results based on matrix-product variational
calculation~\cite{PhysRevA.93.464} or extrapolations from exact diagonalization of finite chains to infinite chains~\cite{PhysRevB.46.8268} are broadly
consistent with our finite chain results based on energy gaps. 

In agreement with results for SSH models with hard-core
bosons~\cite{SSH_Rydberg_19} and in contrast to non-interacting
bosons~\cite{PhysRevLett.110.260405}, we find that topological
degeneracy can survive substantial NNN coupling. This can be related
to the fact that our spin model with alternating stronger and weaker
AFM bonds corresponds to a hard-core bosonic SSH model with
additional inter-site density-density interactions. While this
SSH-like scenario is more robust w.r.t FM NNN coupling, AFM NNN
couplings are the one aspect where the Haldane-like scenario offers
more protection. Once $z$-axis anisotropy and NNN couplings are both
active, dimerized SSH-like chains are again more robust than Haldane chains.

We have thus theoretically demonstrated, that dimerized spin chains 
exhibit quite robust $S=\tfrac{1}{2}$ edge states. 
It remains to be investigated, if these edge states can be observed in experiments, similar to the observation of edge states of the Haldane chain~\cite{Mishra2021}.
If so, alternating chains of 'imperfect' spins,
i.e., with $z$-axis anisotropy and longer-range couplings, could
provide a valuable implementation of a protected $S=\tfrac{1}{2}$. 

Edge states of topologically nontrivial spin systems have become an
area of intense research, especially in two dimensions, e.g., for
the Kitaev
model~\cite{PhysRevB.102.134423,PhysRevLett.126.127201,PhysRevLett.125.267206}. In
contrast to edge states of superconducting Kitaev chains, which can be
observed in an STM~\cite{Rev_edges_STM_21}, they are not charged,
which makes them more challenging to investigate. This comes in
addition to the fact that there is currently no clear route to the
implementation of a Kitaev spin liquid on a surface.
A new idea in this direction is the recent proposal to realize the  Kitaev honeycomb model with quantum dots~\cite{cookmeyer2023engineering}.

Our one-dimensional model is based on available building blocks, as the examination and manipulation of adatoms with an STM is a widely used method~\cite{doi:10.1126/science.1082857,doi:10.1126/science.1146110,10.1038/nphys2299,PhysRevLett.111.127203,RevModPhys.91.041001,DELGADO201740,RevModPhys.81.1495,nphys3722,doi:10.1126/science.1214131,doi:10.1021/acs.nanolett.7b02882,GAUYACQ201263,doi:10.1126/science.1125398,10.1038/s41467-019-10103-5,10.1038/nnano.2010.64,doi:10.1126/science.1201725,PhysRevLett.102.256802} and the reviews~\cite{RevModPhys.81.1495,RevModPhys.91.041001,GAUYACQ201263} provide an overview of this topic.
Such finite and imperfect spin chains  may
thus provide a feasible starting point for the observation of
topologically protected edge states in microscopically assembled spin structures.
This has been already exploited to measure edge states of the Haldane chain~\cite{Mishra2021}, where the splitting can be reduced by making the chain
longer. 
Splitting in the dimer scenario can be
tuned by the ratio $J_1$/$J_2$, so that tunneling times can
potentially be made shorter or longer by placing the dimers closer
together or farther apart.

\begin{acknowledgments}
The authors acknowledge support by the state of Baden-Württemberg through bwHPC.
\end{acknowledgments}

\section{References}
\bibliography{chains}
\end{document}